\documentclass[british,nofootinbib]{revtex4-2}
\usepackage[T1]{fontenc}
\usepackage[latin9]{inputenc}
\usepackage{geometry}
\usepackage{amsmath}
\usepackage{amssymb}
\usepackage{graphicx}
\usepackage{wasysym}
\usepackage{esint}

\makeatletter

\newcommand{\lyxdot}{.}

\usepackage{color}
\usepackage{babel}

\makeatother

\usepackage{babel}
\begin{document}
\global\long\def\sc{\varphi}%
\global\long\def\sa{\Phi}%

\global\long\def\esp{\mathrm{SBP}}%
\global\long\def\e{\mathrm{end}}%
\global\long\def\nad{\mathrm{nad}}%
\global\long\def\mpl{m_{\mathrm{pl}}}%
\global\long\def\nl{\mathrm{nl}}%
\global\long\def\rnl{r_{\nl}}%
\global\long\def\eff{\mathrm{eff}}%
\global\long\def\vac{\mathrm{vac}}%
\global\long\def\i{\mathrm{i}}%
\global\long\def\pp{++}%
\global\long\def\mm{--}%

\title{Quintessential Inflation and Non-Linear Effects of the Tachyonic Trap
Mechanism}
\author{Mindaugas Kar\v{c}iauskas}
\address{Departamento de Física Teórica and Instituto de Física de Partículas
y del Cosmos IPARCOS}
\address{Universidad Complutense de Madrid, E-28040 Madrid, Spain}
\author{Stanislav Rusak}
\address{Departamento de Física Teórica and Instituto de Física de Partículas
y del Cosmos IPARCOS}
\address{Universidad Complutense de Madrid, E-28040 Madrid, Spain}
\author{Alejandro Saez}
\address{Department of Theoretical Physics, Universidad Autónoma de Madrid,
28049 Madrid, Spain}
\address{Instituto de Física Teórica UAM-CSIC, c/ Nicolás Cabrera 13-15, Universidad
Autónoma de Madrid, 28049 Madrid, Spain}
\begin{abstract}
With the help of the tachyonic trapping mechanism one can potentially
solve a number of problems affecting quintessential inflation models.
In this mechanism we introduce a trapping field with a spontaneous
symmetry breaking potential. When the quintessential inflaton passes
the critical point, a sudden burst of particle production is able
to reheat the universe and trap the inflaton away from the minimum
of its potential. However, self-interactions of the trapping field
suppress particle production and reduce the efficiency of this process.
We develop a method to compute the magnitude of the suppression and
explore the parameter space in which the mechanism can be applied
effectively. 
\end{abstract}
\maketitle

\section{Introduction}

The origin of the accelerated expansion of the Universe remains one
of the biggest puzzles in cosmology. The most direct explanation of
such an expansion is the cosmological constant \citep{Weinberg:1988cp}.
Such an explanation is also perfectly consistent with current observations
\citep{Planck:2018vyg}. Unfortunately, the cosmological constant
explanation suffers from serious theoretical issues. Specifically,
it is difficult to find a theoretical justification for its extremely
small value.

It is thought that explaining the absence of the vacuum energy by
invoking some unknown symmetry should be easier than explaining its
tiny value required to fit the observations. If that is true, the
current accelerated expansion could be driven by the potential energy
of some slow rolling scalar field, called quintessence \citep{Weiss:1987xa,Peebles:1987ek,Wetterich:1987fm,Ratra:1987rm,Ferreira:1997hj,Caldwell:1997ii}.
This mechanism is inspired by cosmic inflation and shares some of
its features. Going one step further, it is argued that unifying the
inflaton and quintessence into one and the same field has additional
benefits. Such models are called quintessential inflation \citep{Peebles:1998qn,Peloso:1999dm,Dimopoulos:2001ix}
(see ref.~\citep{Hossain:2014zma,Geng:2015fla,deHaro:2021swo,Bettoni:2021qfs}
for more recent reviews and references). Apart from being economic
in its field content, another benefit of such models is that inflation
provides initial conditions for the quintessential phase of the evolution.
Otherwise initial conditions are free parameters.

Although, the quintessence and quintessential inflation were introduced
to explain the apparent fine tuning of the cosmological constant,
they bring another set of ``tunings''. Many of those stem from the
fact that the scalar field has to be slowly rolling down the potential
to provide dark energy. This can be achieved if the dynamics of quintessence
is determined by the Hubble friction, which requires that the effective
mass of the field is much smaller that the Hubble parameter today,
$m_{\sc}\equiv\sqrt{V''\left(\sc\right)/2}<H_{0}$, where $H_{0}\simeq10^{-33}\:\mathrm{eV}$.
A scalar field with such a tiny mass is problematic form the observational
as well as theoretical points of view. On the one hand, a generic
scalar field with such a small mass should have been detected, as
it leads to several observable phenomena related to the 5th force
in Nature \citep{Carroll:1998zi}. From the theoretical point of view,
it is difficult to explain such a tiny mass within the framework of
effective field theories \citep{Kolda:1998wq}.

An additional challenge for quintessential inflation models are super-Planckian
field values. In order to explain inflation as well as the dark energy,
which require hugely different energy scales, the potential must have
a very large gradient. Due to such a gradient the $\sc$ field picks
up very large kinetic energy after inflation. The latter is difficult
to dissipate before the field reaches super-Planckian values. This
can be problematic within the effective field theory approach as one
can find it complicated to justify the absence of non-renormalisable
terms in the action \citep{Kolda:1998wq} (models based on $\alpha$-attractors
may avoid this problem, see for example ref.~\citep{Linder:2015qxa,Dimopoulos:2017zvq,AresteSalo:2021wgb}).

In this work we investigate a model first proposed in ref.~\citep{Dimopoulos:2019ogl},
which could address the above mentioned issues. In that work it was
suggested that the quintessence field $\sc$ interacts with another
field $\chi$, which is initially heavy and lies at the origin. As
$\sc$ rolls down its classical potential it passes through a Symmetry
Breaking Point (SBP). At that point the effective mass squared $\chi$
switches from a large positive to a negative value. This triggers
an explosive production of $\chi$ particles. Due to their interactions,
these particles create a very steep quantum potential for the quintessence,
which halts the run of $\sc$ almost instantaneously and traps it
at SBP. We therefore often refer to the $\chi$ field as the ``trapping
field''. The trapping mechanism is inspired by ref.~\citep{Kofman:2004yc}.
However, instead of $\chi$ becoming massless, as suggested in that
work, in our model we consider the mass of $\chi$ to be tachyonic
at SBP.

An attractive feature of this scenario is that conclusions are insensitive
to the curvature of the potential. Barring the arguments of naturalness,
we don't need to impose any constraints on the flatness of the potential
at low energies. Only the height of the potential at SBP is crucial.
Therefore, quantum corrections do not have such a strong negative
impact on this scenario as compared to the conventional quintessence
models. Moreover, as the quintessence field can be trapped before
it reaches the Planck scale, the effects of Planck-suppressed non-renormalisable
terms can also be neglected. Another benefit of this scenario is that
the particle production during the trapping phase can be responsible
for reheating the universe. As it is well known, reheating can be
a challenging issue for generic quintessential inflation models \citep{Dimopoulos:2001ix}.
Finally, at the minimum of the potential both fields, $\sc$ and $\chi$,
are heavy, avoiding the above mentioned 5th force problems.\footnote{One could also consider adding a new challenge to quintessence (and
therefore to quintessential inflation) scenarios from recent measurements
of the Hubble constant $H_{0}$. The value of $H_{0}$, as determined
from the CMB measurements \citep{Planck:2018vyg}, is lower than the
one resulting from local determinations \citep{Freedman:2019jwv,Pesce:2020xfe},
leading to the so called ``Hubble tension'' \citep{Verde:2019ivm}.
Simple quintessence scenarios seem to exacerbate this tension according
to ref.~\citep{Banerjee:2020xcn}. Therefore, if the tension is eventually
confirmed, this will make simple quintessence scenarios with the equation
of state $w>-1$ disfavoured. The scenario discussed in the current
work does not suffer from such a problem because both fields are eventually
trapped, which renders them non-dynamical.}

The fact that the trapping field has a negative mass squared at SBP
brings some technical difficulties. To make the potential bounded
from bellow, we must include a self-interaction term $\lambda\chi^{4}$.
Such an interaction term makes the mode functions of the $\chi$ fields
evolve non-linearly, which in turn affects the particle production
and therefore the trapping (as well as reheating) efficiency. This
is an important difference between the tachyonic trapping mechanism
studied in this work and the one proposed in ref.~\citep{Kofman:2004yc}.\footnote{A variation on the particle production mechanism studied in ref.~\citep{Kofman:2004yc}
is also widely used to reheat the universe in quintessential inflation
models \citep{Campos:2002yk,Agarwal:2017wxo,Dimopoulos:2017tud}.
It is called the ``instant reheating'' mechanism.} In ref.~\citep{Dimopoulos:2019ogl} it was noted that such interactions
can be modeled as an additional contribution to the effective mass
of the trapping field. If that contribution is too large, the $\chi$
field becomes too heavy to be excited, reducing the efficiency of
the trapping and the total energy density stored in $\chi$ particles.

In this work we study the effects of the non-linear evolution more
carefully. We develop analytical methods for computing them and compare
with the numerical simulations, showing a very good agreement. The
former method makes it possible to scan a large space of parameter
values and find regions where the trapping is efficient.

The paper is organised as follows: in section \ref{sec:Trapping}
we introduce and motivate our model and discuss its basic features.
The computation of non-linear effects is done in several steps. First,
in section \ref{sec:no-non-lin} we discuss the particle production
ignoring the self-interaction term. The latter is accounted for by
two different methods. If the suppression of particle production is
small, we can take them into account perturbatively, as it is demonstrated
in section \ref{sec:weak-nonl}. In the opposite regime, where non-linear
effects are very strong, the resonantly produced particles cause ``non-linear
blocking'', whereby newly created particles terminate any further
production. This is discussed in section \ref{sec:nl-blocking}. In
section \ref{sec:prm-spc} we utilise the developed method to explore
the space of parameter values that lead to efficient trapping and
conclude in section \ref{sec:Conclusions}. Finally in the Appendix
we collect some mathematical formulas and more technical derivation
steps.

In this work we use natural units such that $c=\hbar=1$ and the reduced
Planck mass is $\mpl=\left(8\pi G\right)^{-1/2}\simeq2.44\times10^{18}\:\mathrm{GeV}$.

\section{The Model \label{sec:Trapping}}

Consider a simple toy model, which contains two scalar fields: a real
scalar field $\sc$ and a complex scalar field $\Psi$. We write the
Lagrangian of the model as 
\begin{equation}
\mathcal{L}=-\frac{1}{2}\partial_{\mu}\sc\partial^{\mu}\sc-\frac{1}{2}\partial_{\mu}\Psi\partial^{\mu}\Psi^{*}-V\left(\sc,\left|\Psi\right|\right)\,,
\end{equation}
where the potential $V\left(\sc,\left|\Psi\right|\right)$ is of the
form 
\begin{equation}
V\left(\sc,\left|\Psi\right|\right)=V\left(\sc\right)+\frac{1}{4}\lambda\left(\Psi\Psi^{*}-f^{2}\right)^{2}+\frac{1}{2}g^{2}\Psi\Psi^{*}\left(\sc-\sc_{\esp}\right)^{2}\,.\label{Vgen}
\end{equation}
Generically, quintessential inflation models require the potential
$V\left(\sc\right)$ of the scalar field to feature two very flat
plateaus at vastly different energy scales \citep{Dimopoulos:2001ix}.
These flatness conditions allow the potential energy of the field
to dominate over the kinetic one, which is needed to drive the accelerated
expansion of the universe. The plateau at high energies is supposed
to provide a (quasi-)exponential expansion during inflation and the
second plateau should be responsible for the accelerated expansion
of the late universe. To accommodate for such a huge difference in
energy scales at both epochs, the gradient of $V\left(\sc\right)$
between the plateaus must be very large. Therefore, typically, after
the end of inflation, the universe enters into the period of kination,
where the energy budget is dominated by the kinetic energy of the
scalar field \citep{Spokoiny:1993kt,Joyce:1996cp}.

One concrete realisation of such a potential in the context of $\alpha$-attractors
was suggested in ref.~\citep{Dimopoulos:2019ogl}, where the mechanism
of the tachyonic trap is also introduced. In the current work we are
not interested in the detailed shape of $V\left(\sc\right)$. The
only assumptions we make about this part of the potential is that
it contains a plateau at high energy scales to provide inflation and
a very large gradient afterwards, in order to accommodate very small
dark energy scale. In contrast to typical quintessential inflation
scenarios, tachyonic trapping mechanism allows us to dispense with
the second, low energy plateau. Indeed, the only requirement for $V\left(\sc\right)$
at low energies is the height of the potential is $V_{\esp}\equiv V\left(\sc_{\esp}\right)\simeq V_{\mathrm{vac}}\simeq10^{-12}\,\mathrm{eV}^{-4}$,
the exact shape being immaterial. This is a very important benefit
of this scenario, as we don't need to worry about radiative corrections,
which could otherwise spoil attractive quintessence models \citep{Campos:2002yk}.

The second term in eq.~(\ref{Vgen}) is the spontaneous symmetry
breaking potential of $\Psi$, where $f$ is the symmetry breaking
scale. And the last term in the potential specifies interactions between
$\sc$ and $\Psi$. To see the effects of such an interaction let
us decompose $\Psi$ into its radial and angular components as $\Psi\equiv\chi\mathrm{e}^{\i\theta/\mpl}$.
We can thus clearly see that the $\chi$ field is heavy for $\left|\sc\right|\gg\left|\sc_{\esp}\right|$
and therefore anchored at $\left\langle \chi\right\rangle =0$. The
angular component $\theta$ is mainly neglected in this paper, apart
from a few brief comments later.

As the $\sc$ field runs towards $\sc_{\esp}$, its stabilising effect
onto $\chi$ disappears and the $U\left(1\right)$ symmetry is spontaneously
broken. Hence the name ``Symmetry Breaking Point'' (SBP). But before
this happens the whole sequence of events take place, some of which
is the main subject of the current work.

To simplify the discussion we will rescale the $\sc$ field as 
\begin{equation}
\sc\rightarrow\sc-\sc_{\esp}
\end{equation}
without loosing generality. Therefore, neglecting the angular component
$\theta$, we can write our working potential as 
\begin{equation}
V\left(\sc,\chi\right)=V\left(\sc\right)+\frac{1}{2}g^{2}\chi^{2}\sc^{2}+\frac{1}{4}\lambda\left(\chi^{2}-f^{2}\right)^{2}\,,\label{V}
\end{equation}
which we use to model the processes close to $\sc\simeq\text{\ensuremath{\sc}}_{\esp}=0$.

As discussed above, due to the large gradient of $V\left(\sc\right)$,
the universe is assumed to be dominated by the kinetic energy of the
$\sc$ field after inflation up until the first passage of SBP. We
might consider this as an additional motivation for unifying inflation
with dark energy models, as opposed to pure quintessence models, within
the context of the tachyonic trapping mechanism: inflation provides
the necessary conditions required for an effective trapping and reheating
of the universe.

During kination the field is oblivious of its potential and the field's
homogeneous component is governed by a simple equation of motion 
\begin{equation}
\ddot{\sc}+3H\dot{\sc}=0\,,\label{eom}
\end{equation}
where $H$ is the Hubble parameter. In this regime $H$ is given by
\begin{eqnarray}
H & \simeq & \frac{\left|\dot{\sc}\right|}{\sqrt{6}\mpl}\,.\label{Hkin}
\end{eqnarray}
It is easy to integrate eq.~(\ref{eom}) and find 
\begin{eqnarray}
\sc & = & \sqrt{\frac{2}{3}}\mathrm{sign}\left(v\right)\ln\left(1+\sqrt{\frac{3}{2}}\frac{\left|v\right|t}{\mpl}\right)\mpl\,,\label{phi-kin}\\
\dot{\sc} & = & v\mathrm{e}^{-\sqrt{\frac{3}{2}}\frac{\sc/\mpl}{\mathrm{sign}\left(v\right)}}\,.\label{dphi-kin-1}
\end{eqnarray}
In these expressions time is defined such that $t_{\esp}=0$ and 
\begin{equation}
v\equiv\dot{\sc}_{\esp}
\end{equation}
is the field velocity at SBP assuming no particle production. To simplify
the notation, we will take $v>0$ in the rest of the paper.

Initially, for large $\left|\sc\right|$ values, the trapping field
is very heavy and anchored at the origin. The mode functions of $\chi$
satisfy the equation 
\begin{eqnarray}
\ddot{\chi}_{k}+3H\dot{\chi}_{k}+\omega_{k}^{2}\chi_{k} & = & 0\,,\label{chi-EoM}
\end{eqnarray}
where 
\begin{eqnarray}
\omega_{k}^{2} & = & k^{2}+g^{2}\sc^{2}-\lambda f^{2}\,.\label{wk2-def}
\end{eqnarray}
The initial conditions are give by the Bunch-Davies vacuum state,
which is defined by the set of positive frequency modes: 
\begin{eqnarray}
\chi_{k,\mathrm{vac}} & = & \frac{a^{-1}}{\sqrt{2\omega_{k}}}\mathrm{e}^{-\mathrm{i}\intop^{t}\omega_{k}\mathrm{d}t'}\:.\label{chi_vac}
\end{eqnarray}
Such an assumption is justified as for large $g^{2}\sc^{2}$ values
$\chi_{k}$ is heavy and cannot be excited. Interesting processes
start when $\sc$ approaches SBP.

As the field $\sc$ runs close to $\sc_{\esp}=0$, the effective mass
of the $\chi$ field vanishes, as can be seen from the definition
of the potential in eq.~(\ref{V}). Moreover, inevitably for a range
of values $\Delta\sc_{\nad}$ the rate at which that mass changes
becomes non-adiabatic. This leads to the, so called, resonant production
of $\chi$ particles as described in ref.~\citep{Kofman:1997yn}.
Such particles create and effective linear potential for the $\sc$
field, as we demonstrate later. If the production is strong enough,
the linear potential is so steep that it halts the evolution of $\sc$
and anchors it at $\sc_{\esp}$. A very similar process is described
in detail in ref.~\citep{Kofman:2004yc}. In our model, there is
an additional, and in some parameter space dominant, contribution
to the particle production. As the potential has a tachyonic direction
at SBP (see eq.~(\ref{V})) one needs to account for the additional
contribution to the particle production via the process called the
tachyonic resonance \citep{Dufaux:2006ee}.

To properly study the growth of such a linear potential we need to
account for the effects of interaction terms in eq.~(\ref{V}). This
is done by employing the Hartree approximation in our analytical computations
as well as numerical simulations. Effectively this constitutes to
replacing $\chi^{3}\rightarrow3\chi\left\langle \chi^{2}\right\rangle $.
In particular, once such interactions are included, we need to update
eq.~(\ref{eom}) as 
\begin{equation}
\ddot{\sc}+3H\dot{\sc}+g^{2}\left\langle \chi^{2}\right\rangle \sc=0\,,\label{eom-trap}
\end{equation}
where $\left\langle \chi^{2}\right\rangle $ is an expectation value
computed as 
\begin{eqnarray}
\left\langle \chi^{2}\right\rangle \left(t\right) & = & \frac{1}{2\pi^{2}}\intop_{0}^{\infty}k^{2}\left[\left|\chi_{k}\left(t\right)\right|^{2}-\frac{1}{2\left|\omega_{k}\left(t\right)\right|}\right]\mathrm{d}k\,,\label{chi2}
\end{eqnarray}
where the second term is included to subtract one loop contributions
from $\chi$ particles \citep{Kofman:2004yc}.

As one can clearly see from eq.~(\ref{eom-trap}), any $\chi$ particle
production generates an effective potential for the $\sc$ field.
If the production is very efficient, the effective potential becomes
so steep that it halts the run of $\sc$ towards the minimum of $V\left(\sc\right)$
and brings it back towards SBP, where it oscillates with an exponentially
decaying amplitude.

The spontaneous symmetry breaking potential of $\Psi$ contain self-interaction
terms. These terms make the equation of motion of the $\chi$ mode
functions non-linear. Such non-linearities are the main subject of
the study in this work.

To account for self-interactions in the evolution of $\chi_{k}$ we
use the Hartree approximation too. In this case it corresponds to
replacing the non-linear term by an additional (time dependent) contribution
to the mass of the trapping field. That is, the full equation of motion
for $\chi_{k}$ must be written as (c.f. eq.~(\ref{chi-EoM})) 
\begin{eqnarray}
\ddot{\chi}_{k}+3H\dot{\chi}_{k}+\left[k^{2}+g^{2}\sc^{2}-\lambda f^{2}+3\lambda\left\langle \chi^{2}\right\rangle \right]\chi_{k} & = & 0\,.\label{chi-EoM-trap}
\end{eqnarray}
The last two terms make our scenario very different from the one discussed
in ref.~\citep{Kofman:2004yc}, where such terms are absent.

We use eqs.~(\ref{eom-trap})-(\ref{chi-EoM-trap}) together with
the vacuum initial conditions in eq.~(\ref{chi_vac}) to study the
full system numerically. In these simulations we integrate the large
system of coupled differential equations (\ref{chi-EoM-trap}) for
a very broad range of $k$ values. But to make the analytical progress
we utilise a number of approximations. Eventually the results of analytical
computations are compared with the numerical simulations to confirm
their accuracy.

\section{Evolution Neglecting Self-Interactions\label{sec:no-non-lin}}

The first approximation that we can make is to neglect the expansion
of the Universe. Indeed, the particle production is effective over
a very small time interval $\Delta t$ 
\begin{eqnarray}
\frac{v\Delta t}{\mpl} & \ll & 1\,.\label{tcond}
\end{eqnarray}
Expanding eqs.~(\ref{phi-kin}) and (\ref{dphi-kin-1}) in terms
of this small quantity and keeping only the highest order terms we
find 
\begin{eqnarray}
\sc & \simeq & vt\ll\mpl\,,\label{phi-linear}\\
\dot{\sc} & \simeq & v=\mathrm{const}\,,
\end{eqnarray}
which are valid during the first passage of SBP.

On the other hand, applying the condition in eq.~(\ref{tcond}) to
eq.~(\ref{Hkin}) we find 
\begin{eqnarray}
H_{\esp}\Delta t & \ll & 1\,,\label{Hsbp-cond}
\end{eqnarray}
where 
\begin{eqnarray}
H_{\esp} & \simeq & \frac{v}{\sqrt{6}\mpl}\,.\label{Hsbp}
\end{eqnarray}
Therefore, it is safe to neglect the Hubble expansion, i.e. the scale
factor can be set to $a=1$, when analytically computing particle
production during \emph{the first passage} of SBP.\footnote{When solving the equations numerically we do not make this approximation.}

If the the particle production is sufficiently effective, each subsequent
passage of SBP results in more $\chi$ particles being produced. Due
to interactions, these particles backreact onto the motion of $\sc$
and result in an exponential decrease of its oscillation amplitude.
Therefore, the understanding of the first burst of particle production
is important to be able to determine the efficiency of the trapping
mechanism as a whole.

Self-interactions of the $\chi$ field can affect such an efficiency
substantially in some of the parameter space. This can be seen from
eq.~(\ref{chi-EoM-trap}), where the $\left\langle \chi^{2}\right\rangle $
term can be interpreted as an additional, time dependent contribution
to the effective mass of the $\chi$ field. As new particles are produced,
$\left\langle \chi^{2}\right\rangle $ grows rapidly. But the growth
of $\left\langle \chi^{2}\right\rangle $ also suppresses further
particle production. In some cases, the growth of $\left\langle \chi^{2}\right\rangle $
can be so fast that it blocks any further particle production once
it even barely started. We call this effect a ``non-linear blocking''.

However, to properly account for the non-linear effects onto the efficiency
of particle production, and therefore the trapping, we first consider
the case without self-interactions in this section. In the next section,
we include non-linearities ``perturbatively'', if they are small,
or compute the effects of non-linear blocking in section~(\ref{sec:nl-blocking}),
if non-linearities are strong.

In the narrow window of particle production we apply the condition
in eq.~(\ref{tcond}), which also leads to the conditions in eqs.~(\ref{phi-linear})
and (\ref{Hsbp}). Therefore, without the non-linear term, one can
write eq.~(\ref{chi-EoM}) during the first passage of SBP as 
\begin{eqnarray}
\ddot{\chi}_{k}^{\left(0\right)}+\omega_{k}^{2}\chi_{k}^{\left(0\right)} & \simeq & 0\,,\label{chi-EoM-H0}
\end{eqnarray}
where $\omega_{k}^{2}$ is given by 
\begin{eqnarray}
\omega_{k}^{2} & \simeq & k^{2}-\lambda f^{2}+g^{2}v^{2}t^{2}\,.\label{wk2-apx}
\end{eqnarray}

Eq.~(\ref{chi-EoM-H0}) can be solved exactly in terms of Parabolyc
Cylinder Functions (PCF).\footnote{We summarise a few relevant properties of PCF and derive other useful
relations in the appendix, section~\ref{sec:PCF}.} But to make generalisations and the connection to the existing literature
easier, we use the WKB approximate expressions sufficiently far from
the particle production region.\footnote{This region has to be close enough for the condition in eq.~(\ref{tcond})
to be satisfied. However, as we will see later, this requirement is
not restrictive at all.} However, to compute the solutions in the neighbourhood of SBP, i.e.
at $t\simeq0$, the use of PCF is essential.

A similar computation is provided in ref.~\citep{Kofman:1997yn}
where only the parametric particle production is considered. While
the work in ref.~\citep{Dufaux:2006ee} provides a similar computation
but with parameters that make the tachyonic particle production dominant.
In our case, both regimes are relevant. Therefore, we develop a computation
which allows to account for both possibilities simultaneously.

Let us first denote the time $t_{\pm}$ such that 
\begin{eqnarray}
t_{\pm} & = & \begin{cases}
0 & \text{for modes with }k^{2}\le\lambda f^{2}\\
\omega_{k}^{2}\left(t_{\pm}\right)=0 & \text{for modes with }k^{2}>\lambda f^{2}
\end{cases}\,,
\end{eqnarray}
and $\mathrm{sing}\left(t_{\pm}\right)=\pm1$. Then, in the region
$t\ll t_{k-}$ $\omega_{k}^{2}$ changes adiabatically, i.e. $\left|\dot{\omega}_{k}\right|\ll\left|\omega_{k}\right|^{2}$
and $\left|\ddot{\omega}_{k}\right|\ll\left|\omega_{k}\right|^{3}$,
and we can write the WKB solution of eq.~(\ref{chi-EoM-H0}) as 
\begin{eqnarray}
\chi_{k-}\left(t\right) & = & \frac{\alpha_{k-}}{\sqrt{2\omega_{k}}}\mathrm{e}^{-\i\intop_{t_{0}}^{t}\omega_{k}\mathrm{d}t'}+\frac{\beta_{k-}}{\sqrt{2\omega_{k}}}\mathrm{e}^{\i\intop_{t_{0}}^{t}\omega_{k}\mathrm{d}t'}\label{chi-WKB-neg}
\end{eqnarray}
where $\chi_{k-}\left(t\right)\equiv\chi_{k}\left(\left|t\right|\gg t_{k-}\right)$.

Long after the first burst of particle production, at $t\gg t_{k+}$,
the rate of change of $\omega_{k}$ is again adiabatic and we can
write 
\begin{eqnarray}
\chi_{k+}\left(t\right) & = & \frac{\alpha_{k+}}{\sqrt{2\omega_{k}}}\mathrm{e}^{-\i\intop_{t_{+}}^{t}\omega_{k}\mathrm{d}t'}+\frac{\beta_{k+}}{\sqrt{2\omega_{k}}}\mathrm{e}^{\i\intop_{t_{+}}^{t}\omega_{k}\mathrm{d}t'}\label{chi-WKB-pos}
\end{eqnarray}
where $\chi_{k+}\left(t\right)\equiv\chi_{k}\left(t\gg t_{k+}\right)$.
In both cases Bogoliubov coefficients are normalised as $\left|\alpha_{k\pm}\right|^{2}-\left|\beta_{k\pm}\right|^{2}=1$.

As was mentioned above, initially the trapping field is heavy and
remains in its vacuum state. This corresponds to choosing the positive
frequency mode of eq.~(\ref{chi-WKB-neg}), i.e. $\alpha_{k-}=1$
and $\beta_{k-}=0$.

To find the connection formulas between coefficients $\alpha_{k-}$,
$\beta_{k-}$ and $\alpha_{k+}$, $\beta_{k+}$ one can use standard
methods employed in quantum mechanics (this is discussed in many textbooks
on quantum mechanics, for example \citep{merzbacher:1998book}).

For modes with the wavenumber $k^{2}>\frac{2gv}{3^{3/2}}+\lambda f^{2}$,
$\omega_{k}^{2}$ is always positive and remains adiabatic. Therefore,
such modes do not undergo any amplification and WKB solutions in eqs.~(\ref{chi-WKB-neg})
and (\ref{chi-WKB-pos}) can be ``connected directly'', i.e. $\alpha_{k-}=\alpha_{k+}$
and $\beta_{k-}=\beta_{k+}$. For modes with smaller $k$ value the
two adiabatic regimes are interrupted by a non-adiabatic regime, in
the case of $\lambda f^{2}<k^{2}<\frac{2gv}{3^{3/2}}+\lambda f^{2}$,
or also by a regime with a negative $\omega_{k}^{2}<0$, in the case
of $k^{2}<\lambda f^{2}$. In those cases the connection between $\alpha_{k\pm}$
and $\beta_{k\pm}$ coefficients is more complicated, which is the
manifestation of the particle production.

When non-adiabaticity is broken, the two adiabatic regions can be
connected using the solutions in terms of PCF 
\begin{eqnarray}
\chi_{k} & = & a_{k}W\left(\kappa,\tau\right)+b_{k}W\left(\kappa,-\tau\right)\,,\label{chi-PCF}
\end{eqnarray}
where we defined 
\begin{eqnarray}
\kappa & \equiv & \frac{\lambda f^{2}-k^{2}}{2gv}\,,\label{kappa-def}\\
\tau & \equiv & \sqrt{2gv}t\,.
\end{eqnarray}
Generically such solutions, in terms of PCF constitute a very good
approximation. But in our case they are exact due to the form of $\omega_{k}$
in eq.~(\ref{wk2-apx}). This being the case, we use the solution
in eq.~(\ref{chi-PCF}) in the whole region where $\omega_{k}^{2}$
is negative to simplify the derivation. Although, we could also make
use of the WKB approximation in the regime where $\omega_{k}^{2}\ll0$,
similarly to what we do in section \ref{sec:nl-blocking}.

To derive the connection formulas, on can be extended eq.~(\ref{chi-PCF})
to regions $\left|t\right|\apprge\left|t_{\pm}\right|$, where the
WKB expressions as well as the expression in eq.~(\ref{chi-PCF})
are both sufficiently good approximations.\footnote{Note, that such a region does not exist in general. But it certainly
does in our case.} We first match the WKB expression in eq.~(\ref{chi-WKB-neg}) with
the one in eq.~(\ref{chi-PCF}). This procedure gives (see eqs.~(\ref{c1-con})
and (\ref{c2-con})) 
\begin{eqnarray}
a_{k} & = & \i\left(\frac{\sqrt{1+\mathrm{e}^{2\pi\kappa}}-\mathrm{e}^{\pi\kappa}}{2\sqrt{2gv}}\right)^{1/2}\mathrm{e}^{-\i\left(\theta_{k}+\varphi_{k}\right)}\,,\label{a-prm}\\
b_{k} & = & \left(\frac{\sqrt{1+\mathrm{e}^{2\pi\kappa}}+\mathrm{e}^{\pi\kappa}}{2\sqrt{2gv}}\right)^{1/2}\mathrm{e}^{-\i\left(\theta_{k}+\varphi_{k}\right)}\,,\label{b-prm}
\end{eqnarray}
where we used vacuum initial conditions at $t_{0}$. $\theta_{k}$
in the above expressions is the phase accumulated from the initial
moment $t_{0}$ to $t_{-}$ 
\begin{eqnarray}
\theta_{k} & \equiv & \intop_{t_{0}}^{t_{-}}\omega_{k}\mathrm{d}t\,,
\end{eqnarray}
and $\varphi_{k}$ is defined in eq.~(\ref{varphik-def}) as 
\begin{eqnarray}
2\varphi_{k} & \equiv & \frac{1}{2}\pi+\mathrm{arg}\left[\Gamma\left(\frac{1}{2}+\i\kappa\right)\right]+\kappa\left(1-\ln\left|\kappa\right|\right)
\end{eqnarray}
To find the final value of $\chi_{k}$, after the resonance is over,
we do the second matching of eq.~(\ref{chi-PCF}) to the WKB solution
in the $t>t_{+}$ region in eq.~(\ref{chi-WKB-pos}). This gives
\begin{eqnarray}
\alpha_{k+} & = & \sqrt{1+\mathrm{e}^{2\pi\kappa}}\mathrm{e}^{-\i\varphi_{k}}\label{alpha-(0)}\\
\beta_{k+} & = & \mathrm{e}^{\pi\kappa}\mathrm{e}^{-2\i\left(\theta_{k}+\frac{\pi}{4}\right)}\label{beta-(0)}
\end{eqnarray}

In summary, after the particle production is over, the mode functions
of the trapping field evolve according to eq.~(\ref{chi-WKB-pos})
with the constants given in eqs.~(\ref{alpha-(0)}) and (\ref{beta-(0)}).

We can use this result to compute the occupation number $n_{k}$ defined
as \citep{Kofman:1997yn} 
\begin{eqnarray}
n_{k} & = & \frac{\left|\omega_{k}\right|}{2}\left[\frac{\left|\dot{\chi}_{k}\right|^{2}}{\left|\omega_{k}\right|^{2}}+\left|\chi_{k}\right|^{2}\right]-\frac{1}{2}\,,\label{nk-def}
\end{eqnarray}
which is approximately constant in the WKB region with $\omega_{k}^{2}>0$.
Plugging in eq.~(\ref{chi-WKB-pos}) into eq.~(\ref{nk-def}) with
$\alpha_{k+}$ and $\beta_{k+}$ provided in eqs.~(\ref{alpha-(0)})
and (\ref{beta-(0)}) we find 
\begin{eqnarray}
n_{k}^{\left(0\right)} & = & \left|\beta_{k+}\right|^{2}=\mathrm{e}^{\pi\frac{\lambda f^{2}-k^{2}}{gv}}\,,\label{nk-(0)}
\end{eqnarray}
where the superscript `$\left(0\right)$' indicates that this quantity
is computed neglecting non-linearities.

Integrating this expression according to 
\begin{eqnarray}
n_{\chi} & = & \frac{1}{2\pi^{2}}\intop_{0}^{\infty}k^{2}n_{k}\mathrm{d}k\label{nchi-def}
\end{eqnarray}
gives the particle number density. At the ``zeroth order'' in non-linearities
this expression leads to 
\begin{eqnarray}
n_{\chi}^{\left(0\right)} & = & \left(\frac{\sqrt{gv}}{2\pi}\right)^{3}\mathrm{e}^{\pi\frac{\lambda f^{2}}{gv}}\,.
\end{eqnarray}

\section{The Effect of Self-Interactions\label{sec:weak-nonl}}

In the previous section we computed the particle number density after
the first passage of SBP neglecting self-interactions of the trapping
field $\chi$. Such interactions introduce non-linear terms in the
equation of motion of $\chi_{k}$, which makes it impossible to find
exact analytical solutions. Unfortunately, in a large parameter space
self-interactions affect the final particle number density considerably
and cannot be neglected. In this and the next sections we develop
methods to compute the effects of such non-linearities.

First, we are going to employ the Hartree approximation as was already
mentioned in the discussion leading to eq.~(\ref{chi-EoM-trap}).
Such an approximation should be sufficient when the non-linear term
is small. In the opposite regime, when it becomes large, the Hartree
approximation breaks. However, it is reasonable to think that this
does not affect the final results much. The reason being that large
non-linear term blocks any further particle production. Therefore,
we only need to find the evolution of $\left\langle \chi^{2}\right\rangle $
term until just before the non-linear blocking, where we can use the
results of section \ref{sec:no-non-lin}.

In the current section we are going to study the parameter region
where the effects of non-linear evolution can be accounted for perturbatively.
The case of strong non-linearities will be considered in the next
section separately.

Our method consists in solving for $\chi_{k}$ iteratively. At the
zeroth order we take the solution derived in the previous section,
which provides us with the method to compute $\left\langle \chi^{2}\right\rangle ^{\left(0\right)}\left(t\right)$.
At the next order we solve the equation (c.f. eqs.~(\ref{chi-EoM-H0})
and (\ref{wk2-apx})) 
\begin{eqnarray}
\ddot{\chi}_{k}^{\left(1\right)}+\left(\omega_{k}^{2}+\delta m^{2}\right)\chi_{k}^{\left(1\right)} & = & 0\,.
\end{eqnarray}
Notice, that at this order we included an additional contribution
to the effective mass squared 
\begin{eqnarray}
\delta m^{2} & \equiv & 3\lambda\left\langle \chi^{2}\right\rangle _{0}^{\left(0\right)}=\mathrm{constant}\,,
\end{eqnarray}
where $\left\langle \chi^{2}\right\rangle _{0}^{\left(0\right)}\equiv\left\langle \chi^{2}\right\rangle ^{\left(0\right)}\left(t=0\right)$
is the expectation value of $\chi^{2}$ computed at zeroth order and
evaluated at the time $t=0$, i.e. at SBP.

For modes that satisfy $\lambda f^{2}<k^{2}<\frac{2gv}{3^{3/2}}+\lambda f^{2}$
we cannot use the WKB solutions in the neighbourhood of $t=0$ as
$\omega_{k}^{2}$ does not evolve adiabatically in that region. Thus
we will use the exact solutions in terms of PCF in eq.~(\ref{chi-PCF}).
For modes with $k^{2}\ll\lambda f^{2}$ WKB approximation does give
a good solution at $t\simeq0$. However, to simplify the argument
and provide a unified framework we will also use the solution in eq.~(\ref{chi-PCF}).

Plugging eqs.~(\ref{a-prm}) and (\ref{b-prm}) into eqs.~(\ref{chi-PCF})
and (\ref{chi2}) we find 
\begin{eqnarray}
\left\langle \chi^{2}\right\rangle _{0}^{\left(0\right)} & = & \frac{1}{2\pi^{2}\sqrt{2gv}}\intop_{0}^{\infty}k^{2}\left[W^{2}\left(\kappa,0\right)\sqrt{1+\mathrm{e}^{2\pi\kappa}}-\frac{1}{2\sqrt{\kappa}}\right]\mathrm{d}k\,.\label{chi2-intm}
\end{eqnarray}
The above integrand peaks at some $k_{*}$ value. For the most of
the parameter space this value is such that $\left|\kappa_{*}\right|>1/2$,
where $\kappa\left(k\right)$ is defined in eq.~(\ref{kappa-def}).
This fact justifies the usage of an approximate value of $W\left(\kappa,0\right)$
in eq.~(\ref{W-kappa-large}) 
\begin{eqnarray}
W^{2}\left(\kappa,0\right) & \simeq & \frac{1}{2\sqrt{\left|\kappa\right|}}\,.\label{large-kappa-apx}
\end{eqnarray}
Plugging it into eq.~(\ref{chi2-intm}), we obtain 
\begin{eqnarray}
\left\langle \chi^{2}\right\rangle ^{\left(0\right)}\left(t=0\right) & \simeq & \frac{1}{4\pi^{2}}\intop_{0}^{\infty}k^{2}\frac{\sqrt{1+\mathrm{e}^{2\pi\kappa}}-1}{\sqrt{\left|k^{2}-\lambda f^{2}\right|}}\mathrm{d}k\,.\label{chi(0)-apx}
\end{eqnarray}
The integral can be computed using the Laplace's approximation. As
the computation involves a few steps we summarise them in Appendix~\ref{sec:integral}.
The final result depends on the ratio 
\begin{eqnarray}
Q & \equiv & \pi\frac{\lambda f^{2}}{gv}\,.\label{Q-def}
\end{eqnarray}
For the $Q$ values in the range $0<Q<\text{a few}$, the largest
contribution to the integral comes from the mode (see eq.~(\ref{xst-small}))
\begin{eqnarray}
k_{*p}^{2} & \simeq & \frac{gv}{\pi}\,.
\end{eqnarray}
In this case the approximate value of eq.~(\ref{chi(0)-apx}) can
be computed to be 
\begin{eqnarray}
\left\langle \chi^{2}\right\rangle _{0p}^{\left(0\right)} & \simeq & \frac{gv}{\left(2\pi\right)^{3}}\sqrt{\frac{\pi/2}{\left|1-Q\right|}}\mathrm{e}^{Q-1}\,.\label{chi2(0)0-p}
\end{eqnarray}

In the opposite regime, with very large $Q$, the biggest contribution
to the integral comes from the modes (see eq.~(\ref{xst-large}))
\begin{eqnarray}
k_{*t}^{2} & \simeq & \frac{2gv}{\pi}\,.\label{k_star_t}
\end{eqnarray}
Thus this is the regime were particle production is overwhelmingly
dominated by the tachyonic particle production ($\lambda f^{2}=Qk_{*t}^{2}/2\gg k_{*t}^{2}$).
The expectation value of $\chi^{2}$ at $t=0$ in this case is 
\begin{eqnarray}
\left\langle \chi^{2}\right\rangle _{0t}^{\left(0\right)} & \simeq & \frac{gv}{2\pi^{3}}\sqrt{\frac{\pi/2}{\left|1-\frac{1}{2}Q\right|}}\mathrm{e}^{\frac{1}{2}Q-1}\,,\label{chi2(0)0}
\end{eqnarray}
where we used eq.~(\ref{Iapx-large}).

The $\delta m^{2}\equiv3\lambda\left\langle \chi^{2}\right\rangle _{0}^{\left(0\right)}$
term only adds a positive constant contribution to $\omega_{k}^{2}$.
Therefore, it is easy to deduce that at the 1st order in this approximation
the particle number density can be written as 
\begin{eqnarray}
n_{\chi}^{\left(1\right)} & = & n_{\chi}^{\left(0\right)}\mathrm{e}^{-\pi\frac{3\lambda\left\langle \chi^{2}\right\rangle _{0i}^{\left(0\right)}}{gv}}\label{nchi(1)}
\end{eqnarray}
where $\left\langle \chi^{2}\right\rangle _{0i}^{\left(0\right)}$
is either $\left\langle \chi^{2}\right\rangle _{0p}^{\left(0\right)}$
(for $Q<\text{a few}$) or $\left\langle \chi^{2}\right\rangle _{0t}^{\left(0\right)}$
(for $Q\gg1$).

We compare this result with the numerical simulations in figure~\ref{fig:num-anl-comp}.
The result in eq.~(\ref{nchi(1)}) is shown as the black curves in
the first column of plots. The blue dot-dashed line in that figure
corresponds to $Q=1$. To the left of that line parametric particle
production dominates. While on the right hand side, the tachyonic
particle production dominates.

On the left hand side of the dot-dashed red line in figure~\ref{fig:num-anl-comp}
the self-interaction induced suppression factor is small, and we can
apply the perturbative result in eq.~(\ref{nchi(1)}). However, as
$\left\langle \chi^{2}\right\rangle _{0}^{\left(0\right)}$ becomes
very large, this result is rendered inadequate. We draw the boundary
between the two regions (which is shown by the red line) at $n_{\chi}^{\left(1\right)}/n_{\chi}^{\left(0\right)}=1/2$.
This is equivalent to saying that the perturbative computation is
used in the region that satisfies the condition 
\begin{eqnarray}
\sqrt{Q-1}\mathrm{e}^{1-Q} & < & \frac{3\lambda}{\ln16\left(2\pi\right)^{3/2}}\,.\label{nl-cond}
\end{eqnarray}
In the opposite regime, the non-linear blocking terminates particle
production. We discuss this case in the next section.

\section{The Strongly Non-Linear Regime \label{sec:nl-blocking}}

The tachyonic mass of the $\chi$ field at SBP makes the particle
production much more effective. \emph{A priori} one would expect that
this makes the trapping more efficient. However, to make the potential
bounded from bellow, we need to introduce a self-coupling term. As
we saw in the previous paragraph, this term suppresses the particle
production. If the latter is very efficient, the non-linear blocking
shuts it down completely.

The non-linear blocking happens due to the rapidly increasing $3\lambda\left\langle \chi^{2}\right\rangle $
term. Once this term reaches $3\lambda\left\langle \chi^{2}\right\rangle \simeq\omega_{k}^{2}$,
the trapping field becomes too heavy for further excitations. We determine
the exact proportionality constant from our numerical simulations.
Indeed, we find that particle production is shut off at the moment
$t_{\nl}$, when the condition

\begin{equation}
\omega_{k_{*t}}^{2}\left(t_{\mathrm{nl}}\right)=-c\cdot3\lambda\left\langle \chi^{2}\right\rangle ^{\left(0\right)}\left(t_{\mathrm{nl}}\right)\label{tnl}
\end{equation}
is satisfied, where $c\simeq2.15$.

In principle we can use eqs.~(\ref{chi-PCF}), (\ref{a-prm}) and
(\ref{b-prm}) to find the time evolution of $\left\langle \chi^{2}\right\rangle ^{\left(0\right)}\left(t\right)$,
but this is not very illuminating. Instead, we are going to use the
WKB approximate relations. This is made possible by the fact that
non-linear blocking happens only in the regime of strong tachyonic
instability. This regime corresponds to very large $\kappa$ (defined
in eq.~(\ref{kappa-def})), where PCF $W\left(\kappa,\tau\right)$
can be approximated by their WKB expressions as in eq.~(\ref{y2}).

Let us write the latter as 
\begin{eqnarray}
\chi_{k} & \simeq & \frac{\alpha_{k}}{\sqrt{2\left|\omega_{k}\right|}}\mathrm{e}^{-\intop_{0}^{t}\left|\omega_{k}\right|\mathrm{d}t}+\frac{\beta_{k}}{\sqrt{2\left|\omega_{k}\right|}}\mathrm{e}^{\intop_{0}^{t}\left|\omega_{a}\right|\mathrm{d}t}
\end{eqnarray}
where the constants $\alpha_{k}$, $\beta_{k}$ are normalised as
$\alpha_{k}\beta_{k}^{*}-\alpha_{k}^{*}\beta_{k}=\i$. They can be
related to $a_{k}$, $b_{k}$ in eq.~(\ref{BD}) as 
\begin{eqnarray}
\alpha_{k} & = & \left(8gv\kappa\right)^{1/4}W\left(\kappa,0\right)a_{k}\\
\beta_{k} & = & \left(8gv\kappa\right)^{1/4}W\left(\kappa,0\right)b_{k}
\end{eqnarray}
Plugging this expression into eq.~(\ref{chi2}) we find 
\begin{eqnarray}
\left\langle \chi^{2}\right\rangle ^{\left(0\right)}\left(t\right) & = & \frac{1}{4\pi^{2}}\intop_{0}^{\infty}\frac{k^{2}}{\left|\omega_{k}\right|}\left(\left|\alpha_{k}\right|^{2}\mathrm{e}^{-2\intop_{0}^{t}\left|\omega_{k}\right|\mathrm{d}t}+\left|\beta_{k}\right|^{2}\mathrm{e}^{2\intop_{0}^{t}\left|\omega_{k}\right|\mathrm{d}t}-1\right)\mathrm{d}k\label{chi2-apx}
\end{eqnarray}
where we used the large $\kappa$ approximation of $W\left(\kappa,0\right)$
(eq.~(\ref{W-kappa-large})). Evaluating this integral at $t=0$
eq.~(\ref{chi2(0)0}) is recovered. As the non-linear blocking only
happens in the regime where tachyonic particle production is dominant,
we concentrate on the $k_{*t}\ll\sqrt{\lambda}f$ mode, which is defined
in eq.~(\ref{k_star_t}).

For sufficiently large $\left|t\right|$, the exponentially increasing
term dominates eq.~(\ref{chi2-apx}). Moreover, due to the smallness
of $k_{*t}$, we can approximate $\omega_{k_{*}}\simeq\omega_{k=0}$
and write the expression in eq.~(\ref{chi2-apx}) as 
\begin{eqnarray}
\left\langle \chi^{2}\right\rangle ^{\left(0\right)}\left(t\right) & \simeq & \left\langle \chi^{2}\right\rangle _{0t}^{\left(0\right)}\mathrm{e}^{2X\left(t\right)}\,.\label{chi2-t}
\end{eqnarray}
$X\left(t\right)$ in the above is the integral given by 
\begin{eqnarray}
X\left(t\right)\equiv\intop_{0}^{t}\left|\omega_{k=0}\right|\mathrm{d}t & = & \frac{\sc}{2v}\left|\omega_{k=0}\left(\sc\right)\right|+\frac{\lambda f^{2}}{2gv}\arcsin\frac{g\sc}{\sqrt{\lambda f^{2}}}\,.
\end{eqnarray}

Finally to find the value of $t_{\nl}$ when the non-linear blocking
happens we can plug eq.~(\ref{chi2-t}) into (\ref{tnl}). Unfortunately,
this leads to the transcendental expression 
\begin{eqnarray}
\lambda f^{2}-g^{2}\sc^{2}\left(t_{\nl}\right) & = & 3c\lambda\left\langle \chi^{2}\right\rangle _{0t}^{\left(0\right)}\mathrm{e}^{2X\left(t_{\nl}\right)}\,,\label{tnl-eqn}
\end{eqnarray}
which cannot be solved analytically. It is valid in the regime $\lambda f^{2}>g^{2}\sc^{2}$,
so one would be tempted to expand it in terms of $g\sc/\sqrt{\lambda}f$
and keep only a few lower order terms. We found, however, that this
procedure gives a poor agreement with the numerical simulations. Therefore,
in what follows we use the full expression and solve eq.~(\ref{tnl-eqn})
numerically.

The final particle number density, at large $\sc$ values, can be
computed using the equation \citep{Kofman:1997yn} 
\begin{eqnarray}
\left\langle \chi^{2}\right\rangle \left(t\right) & \simeq & \frac{1}{2\pi^{2}}\intop_{0}^{\infty}k^{2}\frac{n_{k}\left(t\right)}{\left|\omega_{k}\left(t\right)\right|}\mathrm{d}k\,.\label{chi2-nchi}
\end{eqnarray}
Substituting again $\omega_{k_{*}}^{2}\left(t\right)\simeq\omega_{k=0}^{2}\left(t\right)$
in the above expression we can factor out $\omega_{k}$, and write
\begin{eqnarray}
n_{\chi}=n_{\chi}\left(t_{\nl}\right) & \simeq & \left|\omega_{k=0}\left(t_{\nl}\right)\right|\left\langle \chi^{2}\right\rangle \left(t_{\nl}\right)\,.\label{nchi-nl}
\end{eqnarray}

\begin{figure}
\begin{centering}
\includegraphics[scale=0.25]{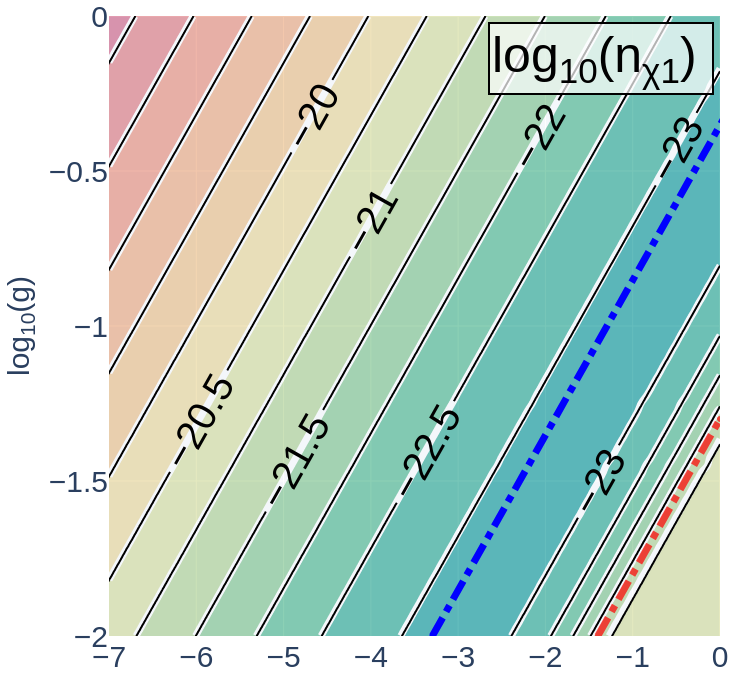}\includegraphics[scale=0.25]{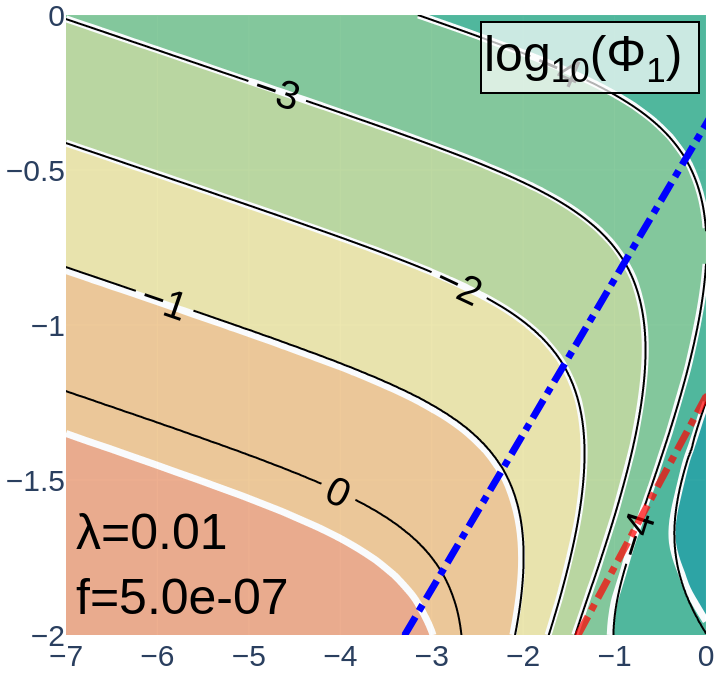} 
\par\end{centering}
\begin{centering}
\includegraphics[scale=0.25]{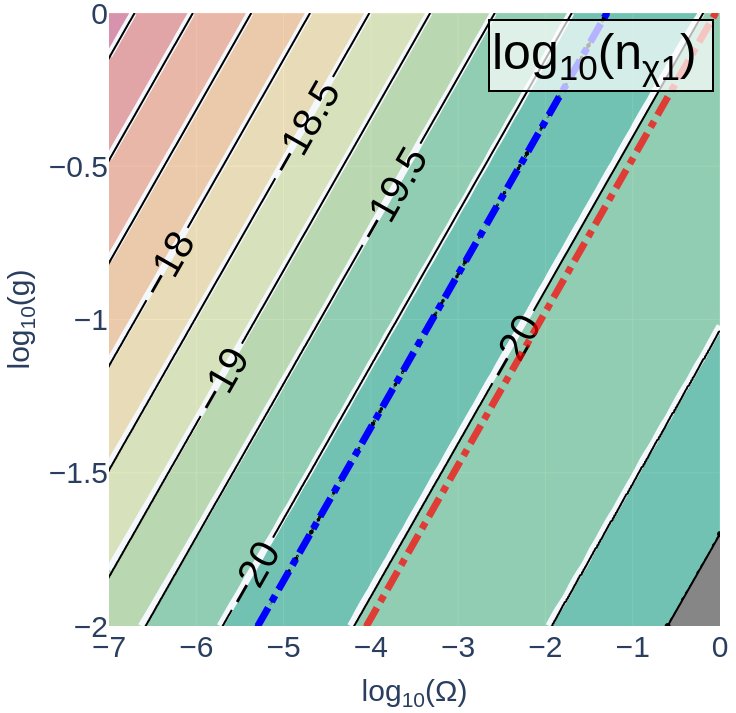}\includegraphics[scale=0.25]{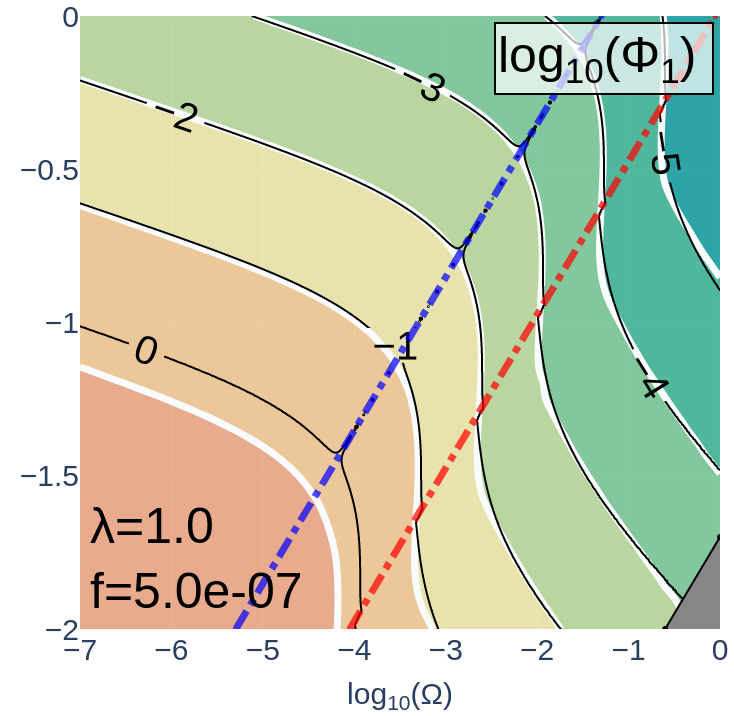} 
\par\end{centering}
\centering{}\caption{\label{fig:num-anl-comp}The comparison of the analytic computations
(solid black curves) with the numerical simulations (white curves
and colour bands). The upper row corresponds to $\lambda=0.01$ and
the lower one to $\lambda=1$. We took $f=5\times10^{-7}$ in both
of these models. On the left column we display the particle number
density $n_{\chi}$ as the function of the interaction strength $g$
and $\Omega\propto v^{-2}$ (see eq.~(\ref{Oapx})). On the right
column we show the amplitude of the first oscillation $\Phi_{1}$
in units of $\protect\mpl$. The blue dot-dashed line corresponds
to $Q=1$ (see eq.~(\ref{Q-def})) and separates the region where
the parametric particle production dominates (left) from the one where
particles are primarily produced by the tachyonic amplification (right).
In the neighbourhood of the blue line the approximation used in eq.~(\ref{large-kappa-apx})
is inadequate. This leads to some artefacts in the plots with large
$\lambda$ values. On the right of the red dot-dashed line the condition
in eq.~(\ref{nl-cond}) is violated, that is, the particle production
is terminated by the non-linear blocking. As one can see from these
plots, the agreement between numerical and analytic computation is
very good. At $\Phi_{1}=\protect\mpl$ the expansion of the universe
starts to be important as can be evidenced from the mismatch between
black and white curves on the right column. The grey triangle in the
bottom-right corner masks the region where our numerical simulations
can no longer solve the equations reliably.}
\end{figure}

This analytic estimate gives a surprisingly good fit to numerical
simulations as can be seen in figure~\ref{fig:num-anl-comp}. The
method developed in this section is applied to computed the right
hand side from the red, dot-dashed line in that figure, which corresponds
to the regime where the condition in eq.~(\ref{nl-cond}) is broken.

\section{The Efficiency of Trapping\label{sec:prm-spc}}

As discussed in section \ref{sec:Trapping}, the interaction between
$\sc$ and $\chi$ fields do not only affect the evolution of the
$\chi$ field but it works the other way round too. The newly created
$\chi$ particles backreact onto the motion of the $\sc$ field as
is demonstrated in eq.~(\ref{eom-trap}). It is clear from that expression
that one can consider such a backreaction as a quantum mechanically
generated effective potential.

Sufficiently late after the first burst of particle production one
can write $\omega_{k_{*}}^{2}\simeq g^{2}\sc^{2}$. Plugging this
approximation into eq.~(\ref{chi2-nchi}) and neglecting the rapidly
oscillating terms (see ref.~\citep{Kofman:1997yn} for details) we
obtain 
\begin{eqnarray}
g^{2}\left\langle \chi^{2}\right\rangle \left(t\right) & \simeq & \frac{gn_{\chi}}{\left|\sc\right|}\,,\label{linV}
\end{eqnarray}
where $n_{\chi}$ is given either in eq.~(\ref{nchi(1)}), for weak
non-linearities, or in eq.~(\ref{nchi-nl}), if particle production
is terminated by the non-linear blocking.

If the particle production is efficient enough and the backreaction
is strong, we can make sure that the $\sc$ field never reaches super-Planckian
values. The need for super-Planckian $\sc$ values in quintessential
inflation models is often recognised as being problematic, as it makes
difficult to justify the absence of non-renormalisable terms in the
original action of the field \citep{Kolda:1998wq}. One can avoid
this problem if the trapping is very strong.

Let us denote the oscillation amplitude of the $\sc$ field as $\Phi$,
and the amplitude after the first burst of particle production as
$\Phi_{1}$. After every other passage of SBP the amplitude decreases.
Therefore it is enough to impose the condition on the oscillation
amplitude after the first passage of SBP 
\begin{eqnarray}
\sa_{1} & < & \mpl\,.\label{Phi1-bound}
\end{eqnarray}

As the Universe expands, newly produced particles are diluted, therefore
reducing the efficiency of the trapping \citep{Brustein:2002mp}.
To prevent this, we require that the time between each burst of particle
production is less than the Hubble time. This condition is much stronger
than the one in eq.~(\ref{Hsbp-cond}). The latter only applies to
the interval of the particle production. Now, we impose a similar
condition to the duration of one oscillation. It is possible to show
that such a condition is equivalent to the one eq.~(\ref{Phi1-bound}).

If the expansion of the universe is neglected, we can write the equation
of motion (\ref{eom-trap}) as 
\begin{eqnarray}
\ddot{\sc}+gn_{\chi}\frac{\sc}{\left|\sc\right|} & \simeq & 0\,,
\end{eqnarray}
where we also used eq.~(\ref{linV}). It is easy to solve this equation
(see ref.~\citep{Kofman:2004yc}). After passing SBP $\sc$ continues
to increase until its initial kinetic energy density is transferred
to the $\chi$ particles. This happens at a time 
\begin{eqnarray}
t_{1} & = & \frac{v}{gn_{\chi1}}\,.\label{t1}
\end{eqnarray}
At that moment the $\sc$ field amplitude is 
\begin{eqnarray}
\sa_{1} & = & \frac{1}{2}\frac{v^{2}}{gn_{\chi1}}\,.\label{Phi1}
\end{eqnarray}
Instead of rolling to the minimum of $V\left(\sc\right)$, as the
classical dynamics would dictate, the $\sc$ field turns around and
runs back towards SBP. At the SBP, the $\chi$ field is approximately
massless again, and $\dot{\sc}\simeq v$.

Plugging eq.~(\ref{t1}) into (\ref{Phi1}) we find 
\begin{eqnarray}
\sa_{1} & = & \frac{1}{2}vt_{1}\,.\label{Phi1t1}
\end{eqnarray}
At the first passage of SBP the universe is dominated by the kinetic
energy of the $\sc$ field, 
\begin{equation}
\rho_{\mathrm{kin}}\equiv\frac{1}{2}v^{2}\label{rkin-def-1}
\end{equation}
and the Hubble parameter is given in eq.~(\ref{Hsbp-cond}). Plugging
this result into eq.~(\ref{Phi1t1}) and using the bound in eq.~(\ref{Phi1-bound})
we find 
\begin{eqnarray}
H_{\esp}t_{1} & < & \sqrt{\frac{2}{3}}\,.\label{eff-cond}
\end{eqnarray}
As one can see, the requirement for sub-planckian field values also
guarantees that the expansion of the universe can be neglected when
computing the particle production during the trapping phase. We confirmed
this using our numerical simulations too, which do include the Hubble
expansion consistently.

One of the goals of this work is to find the parameter range where
the trapping mechanism is efficient, that is, where the condition
in eq.~(\ref{Phi1-bound}), or equivalently in eq.~(\ref{eff-cond}),
is satisfied. To do that we use eq.~(\ref{Phi1}) to scan over the
large space of parameter values, with $n_{\chi1}$ given either by
the expression in eq.~(\ref{nchi(1)}) or (\ref{nchi-nl}) depending
on the value of $Q$.

\begin{figure}
\begin{centering}
\includegraphics[scale=0.5]{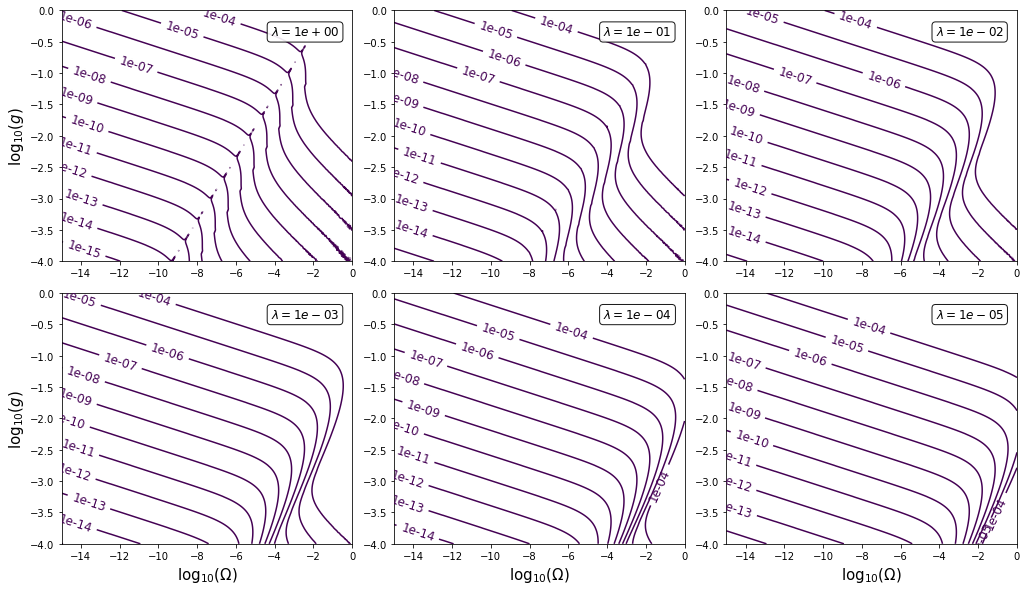} 
\par\end{centering}
\caption{\label{fig:prm-space}The 4-dimensional parameter space $\lambda$,
$f$, $g$ and $v$ for which the trapping mechanism is effective.
Each curve corresponds to a given value of $f$ (shown in the labels)
that results in $\Phi_{1}=\protect\mpl$ (neglecting the expansion
of the universe). Above and to the right of a that curve $\Phi_{1}<\protect\mpl$
for the fixed values of $\lambda$ and $f$. That is, increasing $g$
and/or decreasing $v$, enhances the efficiency of the trap. The definition
of the horizontal axis is shown in eq.~(\ref{O-def}).}
\end{figure}

The scanning is performed over the space of four independent parameters:
$g$, $\lambda$, $f$ and $v$. However, the final constraints should
not be too sensitive to the specifics of the model. The range of the
$\sc$ field values is quite narrow in the window where $\chi$ particles
are produced, and the expression in eq.~(\ref{phi-linear}) should
be a good approximation for a large range of models. For this reason
we express the parameter ranges in figure~\ref{fig:prm-space} in
terms of a physically more relevant quantity: the ratio of the potential
to the kinetic energy at SBP 
\begin{eqnarray}
\Omega & \equiv & \left.\frac{\rho_{\mathrm{pot}}}{\rho_{\mathrm{kin}}}\right|_{\esp}\,.\label{O-def}
\end{eqnarray}

The kinetic energy density is defined in eq.~(\ref{rkin-def-1}).
In our particular model, the potential energy at SBP can be deduced
from eq.~(\ref{V}) 
\begin{equation}
\left.\rho_{\mathrm{pot}}\right|_{\esp}=V_{\mathrm{vac}}+\frac{1}{4}\lambda f^{4}\,,\label{rpot-def}
\end{equation}
where $V_{\mathrm{vac}}\simeq10^{-120}\mpl^{4}\ll\lambda f^{4}$ is
the vacuum energy density. Plugging in eq.~(\ref{rkin-def-1}) we
can therefore write 
\begin{eqnarray}
\Omega & \simeq & \frac{\lambda f^{4}}{2v^{2}}\,.\label{Oapx}
\end{eqnarray}

Let us now consider the possible range of parameter values that we
should scan over. The lowest possible bound on $\Omega$ can be found
by noting that $\rho_{\mathrm{kin}}<V_{*}<\mpl^{4}$. To avoid the
second period of inflation at SBP we also impose the upper bound $\Omega<1$.
Putting these two bounds together one can write 
\begin{eqnarray}
\frac{\lambda f^{4}}{4\mpl^{4}} & <\Omega< & 1\,.\label{O-planck}
\end{eqnarray}

The maximum value of $f$ is constrained from observations. As the
symmetry at SBP is broken one forms a network of cosmic strings. The
tension of such strings is proportional to the symmetry breaking scale
$G\mu=f^{2}/\mpl^{2}$, where $G$ is Newton's constant. The tightest
constraints on $G\mu$ come from CMB measurements, which give $G\mu<10^{-7}$
\citep{Lazanu:2014eya,Lizarraga:2016onn,Ade:2013xla} (see however
ref.~\citep{Bettoni:2019dcw}). In principle $f$ is also bounded
from bellow by the requirement that the symmetry breaking scale is
larger than the scale of the big bang nucleosynthesis (BBN). However,
such a bound is much weaker than the requirement for effective trapping.

The lowest value of $g$ is dictated by the gravitationally induced
interactions. While the upper bounds on $g$ and $\lambda$ are dictated
by the requirement of perturbativity. We will take that value to be
$1$.

Finally, we must consider that before SBP the $\chi$ field must evolve
adiabatically. That is 
\begin{eqnarray}
\left|\frac{\dot{\omega}_{k=0}}{\omega_{k=0}^{2}}\right|_{\sc_{\mathrm{ini}}} & \ll & 1\,,\label{adb-cond-ini}
\end{eqnarray}
where $\sc_{\mathrm{ini}}$ is the field value at the end of inflation
and\footnote{This expression neglects the expansion of the universe. However, we
checked that the results do not change substantially if we include
it.} 
\begin{eqnarray}
\omega_{k=0}^{2} & \simeq & g^{2}\sc^{2}-\lambda f^{2}\,.\label{w2-k0}
\end{eqnarray}
Between $\sc_{\mathrm{ini}}$ and SBP the universe is dominated by
the kinetic energy of the $\sc$ field. Therefore, plugging in eq.~(\ref{dphi-kin-1})
into (\ref{w2-k0}) we find 
\begin{eqnarray}
\left|\frac{\dot{\omega}_{k=0}}{\omega_{k=0}^{2}}\right| & = & \frac{v}{g}\frac{\left|\sc\right|}{\left(\sc^{2}-\frac{\lambda f^{2}}{g^{2}}\right)^{3/2}}\mathrm{e}^{-\sqrt{\frac{3}{2}}\frac{\sc}{\mpl}}\,,
\end{eqnarray}
where $g^{2}\sc^{2}\gg\lambda f^{2}$. This function has a minimum
$\left|\dot{\omega}_{k=0}/\omega_{k=0}^{2}\right|\simeq\frac{v}{g}\frac{3}{8}\mathrm{e}^{2}/\mpl^{2}$
at $\sc\simeq-2\sqrt{\frac{2}{3}}\mpl^{2}$. Imposing the bound in
eq.~(\ref{adb-cond-ini}) we find 
\begin{eqnarray}
\Omega & \gg & \frac{\lambda f^{4}}{g^{2}\mpl^{4}}\,.\label{O-adb-cond}
\end{eqnarray}
The above bound is somewhat stronger than the one in eq.~(\ref{O-planck}),
however we find that it is still much weaker than the bound in eq.~(\ref{Phi1-bound}).

The results of the parameter space scanning are provided in figure~\ref{fig:prm-space}.
In each plot of that figure we draw a curve for a given value of $\lambda$
and $f$ that corresponds to $\Phi_{1}=\mpl$ as computed using eq.~(\ref{Phi1}).
In the space above a given curve, which corresponds to a fixed $\lambda$
and $f$ but larger values of $g$ and/or smaller values of $v$,
the trapping is more efficient and results in a smaller oscillation
amplitude $\Phi_{1}<\mpl$. 

The main goal of the paper is to investigate how non-linearities affect
the oscillation amplitude in the trap. A more detailed study of subsequent
processes falls outside the scope of this work. For completeness,
we only summarise the main points discussed in ref.~\citep{Dimopoulos:2019ogl}.

At $\sc=\Phi_{1}$ the energy density of $\chi$ particles equals
the initial kinetic energy of the $\sc$ field. At that point $\sc$
stops and starts rolling back to SBP. It crosses this point practically
with the same kinetic energy $\dot{\sc}\simeq vt$ \citep{Kofman:2004yc}
and triggers the second burst of particle production. The newly created
particles are added to the total bath, which interact with $\sc$
creating an even steeper potential for the latter. This way the oscillation
amplitude rapidly decreases with each subsequent passage of SBP. The
process continues until one of the two things happen. If the oscillation
amplitude $\Phi$ decays bellow the non-adiabaticity region, the parametric
resonance is no longer effective and particle production stops. Due
to the tachyonic direction in our model, the particle production can
be terminated by a different mechanism. As we saw in section \ref{sec:nl-blocking},
self-interactions can render the $\chi$ field too heavy for any further
excitations. We called this process the ``non-linear blocking''.
In some parameter space it halts any further particle production once
it barely started. Which of the two mechanisms determines the end
of the resonance depends on model parameters.

If the end of particle production is dictated by the first process,
the trapping mechanism can also be responsible for an efficient reheating
of the universe. This is an important advantage of the mechanism,
as reheating in quintessential inflation scenarios is complicated,
due to non-oscillatory potential (see refs.~\citep{Campos:2002yk,Agarwal:2017wxo,Dimopoulos:2017tud,Dimopoulos:2018wfg,Opferkuch:2019zbd,Bettoni:2021zhq,Dimopoulos:2002kt,Dimopoulos:2005bp,Lima:2019yyv}
for other possible alternatives). If, on the other hand, particle
production is terminated by the non-linear blocking, then most likely
reheating is much less efficient. However, this issue requires more
detailed investigation.

Finally, we should mention that in this work we neglected the Goldstone
boson $\theta$ of the $U\left(1\right)$ symmetry breaking. There
are two possibilities, depending on a specific implementation. In
ref.~\citep{Dimopoulos:2019ogl} it was suggested that $U\left(1\right)$
might be a global symmetry. In which case $\theta$ could constitute
the QCD axionic dark matter in the Universe (once the potential of
$\theta$ is lifted) \citep{Ringwald:2012hr,Kawasaki:2013ae,Baer:2014eja,Marsh:2015xka}.
In that case $f$ is limited to the ``classical axion window'',
which is the range $10^{9-12}\:\mathrm{GeV}$ (assuming no fine-tuning
of the misalignment angle). This is a particularly attractive possibility
as it conforms to the main motivation of the quintessential inflation
scenario by being minimal in its field content. If, on the other hand,
we do not want to impose such a requirement and $U\left(1\right)$
is taken to be a local symmetry, then the $\theta$ field could be
eaten by gauge bosons to make them heavy. In both cases, it is possible
to avoid the 5th force problems associated with light scalar fields.

\section{Conclusions and Discussion\label{sec:Conclusions}}

The paradigm of quintessential inflation can be very attractive to
model the evolution of our Universe. The primary appeal of such models
is that they do not introduce additional scalar fields beyond the
inflaton. However, this minimalist approach also introduces certain
difficulties. The main one being a mechanism of reheating the universe
after inflation. In addition, with the usual quintessence models it
shares the problem of the 5th force constraints, the need for the
suppression of radiative corrections and explaining the absence of
non-renormalisable terms. To solve those problems one might need to
invoke other fields after all. However, this does not have to violate
the principle of economy of quintessential inflation models. The additional
field(s) can be the same that are already used in cosmology for different
purposes.

This principle was applied in the model proposed in ref.~\citep{Dimopoulos:2019ogl},
where the additional field is the same as used in QCD axion scenarios.
The latter could explain the whole of dark matter in the Universe
\citep{Baer:2014eja}. If the interaction between the inflaton and
the additional field is of the form in eq.~(\ref{Vgen}), the problems
mentioned above can be overcome. On the one hand, the interaction
of this form induces a very rapid and efficient particle production,
as the kinetic energy dominated inflaton zips through the symmetry
breaking point (SBP). Such particles can stop the inflaton at sub-Planckian
values and reheat the universe. On the other hand, the coupling makes
the inflaton heavy in the vacuum, therefore preventing problems associated
with the constraints from the 5th force experiments. Moreover, this
scenario is not sensitive to the precise form of the potential in
the quintessential tail. Only the hight of the potential is important.

The process of particle production is somewhat similar to the mechanism
proposed in ref.~\citep{Kofman:2004yc}. The crucial difference,
however, is that the trapping field is self-interacting. These interactions
lead to the non-linear evolution and change the efficiency of particle
production and consequently the efficiency of the trapping mechanism
and reheating.

In this work we study in detail how self-interactions affect the trapping.
Within the Hartree approximation we develop the analytic formalism
to account for the suppression of particle production. We also compute
the particle number density in the region of non-linear blocking.
In this region the self-interaction of the trapping field is so strong
that it terminates particle production while the effective mass of
the $\chi$ field is still tachyonic. The analytic results agree very
well with numerical simulations as can be seen in figure~\ref{fig:num-anl-comp}.

Using these analytic methods we can very efficiently explore the parameter
space to compute the constraints. In ref.~\citep{Dimopoulos:2019ogl}
the symmetry breaking scale $f$ was fixed. In this work we do not
impose any bounds of $f$ (apart from observational constraints on
cosmic string tension) and explore the full range of possibilities.
Our results are provided in figure~\ref{fig:prm-space}, where we
show the parameter space for efficient trapping. One can see that
this mechanism is effective for a very wide range of parameter values.

If the tachyonic trapping mechanism is indeed realised in Nature,
dark energy would be indistinguishable from the cosmological constant.
However, this mechanism suggests other potentially observable phenomena,
such as the production of primordial gravitational waves as well as
the formation of cosmic strings \citep{Bettoni:2018pbl}. We leave
the study of such possibilities for future publications. 
\begin{acknowledgments}
The work of M.K. and S.R. is supported by the Communidad de Madrid
``Atracción de Talento investigador'' Grant No. 2017-T1/TIC-5305
and MICINN (Spain) project PID2019-107394GB-I00. A. S. is supported
by MICINN (Spain) grant PGC2018-094857-B-I00 and the Spanish Agencia
Estatal de Investigación through the grant ``IFT Centro de Excelencia
Severo Ochoa'' SEV-2016-0597 and CEX2020-001007-S. 
\end{acknowledgments}

\appendix
%dummy comment inserted by tex2lyx to ensure that this paragraph is not empty

\section{The Parabolic Cylinder Functions\label{sec:PCF}}

\subsection{The Properties of Parabolic Cylinder Functions}

In our analytic computations we make an extensive use of Parabolic
Cylinder Functions (PCF). This is due to the fact that these functions
form a complete set of solutions of eq.~(\ref{chi-EoM-H0}). However,
even if the equation of motion of $\chi$ would be more complicated,
PCF give a good approximation in the regions where the evolution of
$\omega_{k}$ is non-adiabatic. Indeed, PCF are widely used in quantum
mechanics precisely in this context: to find formulas connecting regimes
with WKB approximate solutions. In this section we summarise the main
properties of PCF that we use in the main text.

Consider the equation \citep{abramowitz:book1964} 
\begin{eqnarray}
y''+\omega_{\kappa}^{2}y & = & 0\,,\label{pdf-eq}
\end{eqnarray}
where 
\begin{eqnarray}
\omega_{\kappa}^{2} & \equiv & \frac{1}{4}\tau^{2}-\kappa\label{omega-kappa}
\end{eqnarray}
and primes denote derivatives with respect to the independent variable
$\tau$. A generic solution of this equation can be written in terms
of PCF as 
\begin{eqnarray}
y & = & D_{1}W\left(\kappa,\tau\right)+D_{2}W\left(\kappa,\tau\right)\,,\label{sol-PCF}
\end{eqnarray}
where $D_{1,2}$ are integration constants. Equation (\ref{pdf-eq})
has a few equivalent forms related by redefinitions of $\kappa$ and
$\tau$, but the above one is the easiest to apply to our setup.

At $\tau=0$ PCF reduce to 
\begin{eqnarray}
W\left(\kappa,0\right) & = & 2^{-\frac{3}{4}}\left|\frac{\Gamma\left(\frac{1}{4}+\frac{1}{2}\i\kappa\right)}{\Gamma\left(\frac{3}{4}+\frac{1}{2}\i\kappa\right)}\right|^{\frac{1}{2}}
\end{eqnarray}
and 
\begin{eqnarray}
\frac{\mathrm{d}W\left(\kappa,0\right)}{\mathrm{d}\tau} & = & -\frac{1}{2W\left(\kappa,0\right)}\,,
\end{eqnarray}
where $\Gamma$ is the gamma function. In the limit of large $\kappa$
this equation approaches 
\begin{eqnarray}
W\left(\kappa,0\right) & \overset{\kappa\rightarrow\pm\infty}{\longrightarrow} & \frac{1}{\sqrt{2}\left|\kappa\right|^{1/4}}\,.\label{W-kappa-large}
\end{eqnarray}

To make the connection between PCF and the WKB approximate solutions
we made use of several asymptotic forms of $W\left(\kappa,\tau\right)$.
For $\tau^{2}\gg4\left|\kappa\right|$ the $W$ function can be approximated
at the lowest order in $\tau^{-2}$ by 
\begin{eqnarray}
W\left(\kappa,\tau\right) & \approx & \sqrt{\frac{k}{2\left|\tau\right|}}\left(\mathrm{e}^{-\i\omega}+\mathrm{e}^{\i\omega}\right)\label{W1-apx-large_tau}\\
W\left(\kappa,-\tau\right) & \approx & \frac{\i}{\sqrt{2k\left|\tau\right|}}\left(\mathrm{e}^{-\i\omega}-\mathrm{e}^{\i\omega}\right)\label{W2-apx-large_tau}
\end{eqnarray}
where 
\begin{eqnarray}
k & \equiv & \sqrt{1+\mathrm{e}^{2\pi\kappa}}-\mathrm{e}^{\pi\kappa}\label{k-def}\\
\omega & \equiv & \frac{1}{4}\tau^{2}-\frac{1}{2}\kappa\ln\tau^{2}+\frac{1}{4}\pi+\frac{1}{2}\phi_{2}
\end{eqnarray}
and 
\begin{eqnarray}
\phi_{2} & \equiv & \mathrm{arg}\left[\Gamma\left(\frac{1}{2}+\i\kappa\right)\right]\,.\label{phi2}
\end{eqnarray}
Notice that both functions, $W\left(\kappa,\tau\right)$ and $W\left(\kappa,-\tau\right)$,
are real.

In the opposite regime, where $4\left|\kappa\right|\gg\tau^{2}$ the
$W\left(\kappa,\tau\right)$ function can be approximated by 
\begin{eqnarray}
W\left(\kappa,\tau\right) & = & W\left(\kappa,0\right)\mathrm{e}^{-\sqrt{\kappa}\tau+v_{1}}\,,\label{W1-large_kappa}\\
W\left(\kappa,-\tau\right) & = & W\left(\kappa,0\right)\mathrm{e}^{\sqrt{\kappa}\tau+v_{2}}\,,\label{W2-large_kappa}
\end{eqnarray}
and 
\begin{eqnarray}
v_{1,2} & \simeq & \pm\frac{\frac{2}{3}\left(\frac{1}{2}\tau\right)^{3}}{2\sqrt{\kappa}}+\frac{\left(\frac{1}{2}\tau\right)^{2}}{\left(2\sqrt{\kappa}\right)^{2}}\pm\frac{\frac{1}{2}\tau+\frac{2}{5}\left(\frac{1}{2}\tau\right)^{5}}{\left(2\sqrt{\kappa}\right)^{3}}+\ldots\label{niu12}
\end{eqnarray}

\subsection{Relation to WKB Approximation}

For $\tau^{2}\gg\left|\kappa\right|$ and $\tau<0$ the $\omega_{\kappa}$
in eq.~(\ref{omega-kappa}) changes adiabatically. Therefore, we
can also find the solution of eq.~(\ref{pdf-eq}) using WKB approximation.
Let us write this solution as 
\begin{eqnarray}
y_{1} & \simeq & \frac{A_{1}}{\sqrt{2\omega_{\kappa}}}\mathrm{e}^{-\i\intop_{\tau_{0}}^{\tau}\omega_{\kappa}\mathrm{d}\tau}+\frac{A_{2}}{\sqrt{2\omega_{\kappa}}}\mathrm{e}^{\i\intop_{\tau_{0}}^{\tau}\omega_{\kappa}\mathrm{d}\tau}\,.
\end{eqnarray}
$y_{1}$ is nothing else but the approximate expression of the exact
solution in eq.~(\ref{sol-PCF}). Indeed, using eqs.~(\ref{W1-apx-large_tau})
and (\ref{W2-apx-large_tau}) we can find the connection formulas
for the integration constants as 
\begin{eqnarray}
D_{1} & = & \i\sqrt{\frac{k}{2}}\left[A_{1}\mathrm{e}^{-\i\left(\theta_{\kappa}+\varphi_{\kappa}\right)}-A_{2}\mathrm{e}^{\i\left(\theta_{\kappa}+\varphi_{\kappa}\right)}\right]\,,\label{c1-con}\\
D_{2} & = & \frac{1}{\sqrt{2k}}\left[A_{1}\mathrm{e}^{-\i\left(\theta_{\kappa}+\varphi_{\kappa}\right)}+A_{2}\mathrm{e}^{\i\left(\theta_{\kappa}+\varphi_{\kappa}\right)}\right]\,,\label{c2-con}
\end{eqnarray}
where $k$ is defined in eq.~(\ref{k-def}), $\theta_{\kappa}$ is
the phase accumulated from $\tau_{0}$ to $\tau_{-}$ 
\begin{eqnarray}
\theta_{\kappa} & \equiv & \intop_{\tau_{0}}^{\tau_{-}}\omega_{\kappa}\mathrm{d}x
\end{eqnarray}
and $\tau_{-}$ is such that $\omega_{\kappa}^{2}\left(\tau<\tau_{-}\right)>0$,
that is 
\begin{eqnarray}
\tau_{-} & = & \begin{cases}
0 & \text{if }\kappa\le0\\
-\sqrt{2}\kappa & \text{if }\kappa>0
\end{cases}
\end{eqnarray}
The phase $\varphi_{\kappa}$ is defined as 
\begin{eqnarray}
\varphi_{\kappa} & \equiv & \frac{1}{4}\pi+\frac{1}{2}\phi_{2}+\frac{1}{2}\kappa\left(1-\ln\left|\kappa\right|\right)\,,\label{varphik-def}
\end{eqnarray}
where $\phi_{2}$ is given in eq.~(\ref{phi2}).

In the opposite regime, where $\tau^{2}\gg\left|\kappa\right|$ and
$\tau>0$, we can derive similar connection formulas. Let us write
the approximate WKB solution of the equation (\ref{pdf-eq}) as 
\begin{eqnarray}
y_{3} & \simeq & \frac{C_{1}}{\sqrt{2\omega_{\kappa}}}\mathrm{e}^{-\i\intop_{\tau_{+}}^{\tau}\omega_{\kappa}\mathrm{d}\tau}+\frac{C_{2}}{\sqrt{2\omega_{\kappa}}}\mathrm{e}^{\i\intop_{\tau_{+}}^{\tau}\omega_{\kappa}\mathrm{d}\tau}
\end{eqnarray}
where $\tau_{+}$ is given by 
\begin{eqnarray}
\tau_{+} & = & \begin{cases}
0 & \text{if }\kappa\le0\\
\sqrt{2}\kappa & \text{if }\kappa>0
\end{cases}\,,
\end{eqnarray}
Matching the approximate expression of eq.~(\ref{sol-PCF}) one finds
\begin{eqnarray}
C_{1} & = & \frac{kD_{1}+\i D_{2}}{\sqrt{2k}}\mathrm{e}^{-\i\varphi_{\kappa}}\,,\label{b1-con}\\
C_{2} & = & \frac{kD_{1}-\i D_{2}}{\sqrt{2k}}\mathrm{e}^{\i\varphi_{\kappa}}\,.\label{b2-con}
\end{eqnarray}

Notice that using eqs.~(\ref{c1-con}), (\ref{c2-con}) and (\ref{b1-con}),
(\ref{b2-con}) we can derive connection formulas 
\begin{eqnarray}
\left(\begin{array}{c}
C_{1}\\
C_{2}
\end{array}\right) & = & \i\left(\begin{array}{cc}
\sqrt{1+\mathrm{e}^{2\pi\kappa}}\mathrm{e}^{-\i\left(\theta_{\kappa}+2\varphi_{\kappa}\right)} & \mathrm{e}^{\pi\kappa}\mathrm{e}^{\i\theta_{\kappa}}\\
-\mathrm{e}^{\pi\kappa}\mathrm{e}^{-\i\theta_{\kappa}} & -\sqrt{1+\mathrm{e}^{2\pi\kappa}}\mathrm{e}^{\i\left(\theta_{\kappa}+2\varphi_{\kappa}\right)}
\end{array}\right)\left(\begin{array}{c}
A_{2}\\
A_{2}
\end{array}\right)
\end{eqnarray}

We can also find a WKB approximation of the solution in eq.~(\ref{sol-PCF})
in the limit where $\kappa\gg\tau^{2}$. Let write the approximate
solution of eq.~(\ref{pdf-eq}) in this region as 
\begin{eqnarray}
y_{2} & \simeq & \frac{B_{1}}{\sqrt{2\left|\omega_{\kappa}\right|}}\mathrm{e}^{-\intop_{0}^{\tau}\left|\omega_{\kappa}\right|\mathrm{d}\tau}+\frac{B_{2}}{\sqrt{2\left|\omega_{\kappa}\right|}}\mathrm{e}^{\intop_{0}^{\tau}\left|\omega_{\kappa}\right|\mathrm{d}\tau}\,.\label{y2}
\end{eqnarray}
. We can expand $\intop_{0}^{\tau}\left|\omega_{\kappa}\right|\mathrm{d}\tau$
and $\left|\omega_{\kappa}\right|^{-1/2}$ in terms of $\tau^{2}/4\kappa$
and match to the eq.~(\ref{sol-PCF}) using the series expansion
of PCF in eqs.~(\ref{W1-large_kappa}) and (\ref{W2-large_kappa}).
Note, that the series in eq.~(\ref{niu12}) contain terms that would
correspond to higher order WKB approximation than provided in eq.~(\ref{y2}).
After the matching we find 
\begin{eqnarray}
B_{1,2} & = & \sqrt{2}\kappa^{1/4}W\left(\kappa,0\right)D_{1,2}\label{BD}
\end{eqnarray}

\section{Laplace's Approximation \label{sec:integral}}

To compute the integral in eq.~(\ref{chi(0)-apx}) we used the Laplace's
approximation \citep{butler:2007book}. As the derivation of this
particular integral involves a few steps, we recall them in this section.

The approximation can be summarised as 
\begin{eqnarray}
I=\intop_{0}^{\infty}\frac{\mathrm{e}^{Cg\left(x\right)}}{f\left(x\right)}\mathrm{d}x & \simeq & \sqrt{\frac{2\pi}{C\left|g''\left(x_{*}\right)\right|}}\frac{\mathrm{e}^{Cg\left(x_{*}\right)}}{f\left(x_{*}\right)}\,.\label{stdsc}
\end{eqnarray}
where $x_{*}$ is the value of $x$ such that $g\left(x_{*}\right)=\max\left(g\right)$
and primes denote derivatives with respect to $x$. The equality becomes
exact in the limit $C\rightarrow\infty$.

In our case we need to compute the integral of the form 
\begin{eqnarray}
I & = & \intop_{0}^{\infty}\frac{x^{2}}{f\left(x\right)}\left(\sqrt{1+\mathrm{e}^{Q-x^{2}}}-1\right)\mathrm{d}x\,.\label{int}
\end{eqnarray}
To do that, let us first denote 
\begin{eqnarray}
g\left(x\right) & \equiv & \ln\left[x^{2}\left(\sqrt{1+\mathrm{e}^{Q-x^{2}}}-1\right)\right]\,,\label{gx}
\end{eqnarray}
which has a maximum at 
\begin{eqnarray}
Q & = & \ln\left[\frac{x_{*}^{2}-1}{\left(\frac{1}{2}x_{*}^{2}-1\right)^{2}}\right]+x_{*}^{2}\,.\label{Qxst}
\end{eqnarray}
Unfortunately, we cannot solve this transcendental equation using
elementary functions. But it suffices to notice that $x_{*}^{2}\in\left(1,2\right)$.
In the main text we use an approximation such that for $Q\apprle\mathrm{few}$.
This corresponds to $x_{*}$ value 
\begin{eqnarray}
x_{*}^{2} & \simeq & 1\,.\label{xst-small}
\end{eqnarray}
In this limit the approximate value of the integral is given by 
\begin{eqnarray}
I & \simeq & 2^{-3/2}\sqrt{\pi}\frac{\mathrm{e}^{Q-1}}{f\left(x_{*}\right)}\,.\label{Iapx-small}
\end{eqnarray}
In the opposite regime, for $Q\gg1$, the largest contribution to
the integral comes from $x$ values close to 
\begin{eqnarray}
x_{*}^{2} & \simeq & 2\label{xst-large}
\end{eqnarray}
and the approximate value of the integral is given by 
\begin{eqnarray}
I & \simeq & 2\sqrt{\pi}\frac{\mathrm{e}^{\frac{1}{2}Q-1}}{f\left(x_{*}\right)}\,.\label{Iapx-large}
\end{eqnarray}

\bibliographystyle{aapmrev4-2}
\bibliography{draft-qtrap-v4.bbl}

\end{document}